\documentclass[11pt]{article}
\usepackage{moriond,epsfig}
\def\Journal#1#2#3#4{{#1} {\bf #2}, #3 (#4)}

\def\be{\begin{equation}}
\def\ee{\end{equation}}
\def\bea{\begin{eqnarray}}
\def\eea{\end{eqnarray}}

\begin{document}
\vspace*{4cm}

\title{ CHARGED PARTICLE MULTIPLICITY AND LIMITING FRAGMENTATION IN Au+Au COLLISIONS AT RHIC ENERGIES USING THE PHOBOS DETECTOR }
\author{Rachid Nouicer \\
Physics Department, University of Illinois at Chicago\\
E-mail : rachid.nouicer@bnl.gov\\
for the PHOBOS Collaboration }
\address{
B.B.Back$^1$,
M.D.Baker$^2$,
D.S.Barton$^2$,
R.R.Betts$^6$,
M.Ballintijn$^4$,
A.A.Bickley$^7$,
R.Bindel$^7$,
A.Budzanowski$^3$,
W.Busza$^4$,
A.Carroll$^2$,
M.P.Decowski$^4$,
E.Garcia$^6$,
N.George$^{1,2}$,
K.Gulbrandsen$^4$,
S.Gushue$^2$,
C.Halliwell$^6$,
J.Hamblen$^8$,
G.A.Heintzelman$^2$,
C.Henderson$^4$,
D.J.Hofman$^6$,
R.S.Hollis$^6$,
R.Ho\l y\'{n}ski$^3$,
B.Holzman$^2$,
A.Iordanova$^6$,
E.Johnson$^8$,
J.L.Kane$^4$,
J.Katzy$^{4,6}$,
N.Khan$^8$,
W.Kucewicz$^6$,
P.Kulinich$^4$,
C.M.Kuo$^5$,
W.T.Lin$^5$,
S.Manly$^8$,
D.McLeod$^6$,
J.Micha\l owski$^3$,
A.C.Mignerey$^7$,
R.Nouicer$^6$,
A.Olszewski$^{2,3}$,
R.Pak$^2$,
I.C.Park$^8$,
H.Pernegger$^4$,
C.Reed$^4$,
L.P.Remsberg$^2$,
M.Reuter$^6$,
C.Roland$^4$,
G.Roland$^4$,
L.Rosenberg$^4$,
J.Sagerer$^6$,
P.Sarin$^4$,
P.Sawicki$^3$,
W.Skulski$^8$,
S.G.Steadman$^4$,
P.Steinberg$^2$,
G.S.F.Stephans$^4$,
M.Stodulski$^3$,
A.Sukhanov$^2$,
J.-L.Tang$^5$,
R.Teng$^8$,
A.Trzupek$^3$,
C.Vale$^4$,
G.J.van~Nieuwenhuizen$^4$,
R.Verdier$^4$,
B.Wadsworth$^4$,
F.L.H.Wolfs$^8$,
B.Wosiek$^3$,
K.Wo\'{z}niak$^3$,
A.H.Wuosmaa$^1$,
B.Wys\l ouch$^4$\\
$^1$~Argonne National Laboratory, Argonne, IL 60439-4843, USA, 
 $^2$~Brookhaven National Laboratory, Upton, NY 11973-5000, USA,
 $^3$~Institute of Nuclear Physics, Krak\'{o}w, Poland,
 $^4$~Massachusetts Institute of Technology, Cambridge, MA 02139-4307, USA,
 $^5$~National Central University, Chung-Li, Taiwan,
 $^6$~University of Illinois at Chicago, Chicago, IL 60607-7059, USA,
 $^7$~University of Maryland, College Park, MD 20742, USA,
 $^8$~University of Rochester, Rochester, NY 14627, USA
}
\vspace*{-0.5cm}
\maketitle\abstracts{The first measurements of charged particle
pseudorapidity distributions obtained from ${\rm Au + Au}$ collisions at the maximum RHIC energy
(${\rm \sqrt{s_{NN}} = 200}$ GeV) using the PHOBOS detector are presented. A comparison of
the pseudorapidity distributions at energies 130 and 200
GeV for different centrality bins is made, including an estimate of the 
total number of charged particles. Away from the mid-rapidity region, a
comparison between ${\rm Pb + Pb }$ at SPS energy ${\rm
\sqrt{s_{NN}} = }$ 17.3 GeV and ${\rm Au + Au }$ at RHIC energy ${\rm \sqrt{s_{NN}} = }$
130 GeV indicates that the extent of the limiting fragmentation region
grows by about 1.5 units of ${\rm \eta -y_{beam}}$ over this energy range. We also observe 
that the extent of the limiting
fragmentation region is independent of centrality at the same energy,
but that the particle production per participant in the limiting fragmentation region grows at high ${\rm \eta -
y_{beam} \geq -1.5 } $ for more peripheral collisions. In combination with
results from lower energies and from ${\rm \bar{p} + p}$ collisions, these
data permit a systematic analysis of particle production mechanisms in
nucleus-nucleus collisions.         
}
\section{Introduction}\label{sec:intro}
\vspace*{-0.2cm}
Quantum Chromodynamics (QCD) is believed to be the fundamental theory
for strong interactions. According to this theory, hadronic matter under extreme dense and
hot conditions must go through a phase transition~\cite{Col1} to form a
Quark Gluon Plasma (QGP) in which quarks and gluons are no longer confined
to the size of a hadron. Lattice gauge studies of QCD at finite
temperatures have indeed found such a phase transition, though the
exact nature of the transition has not yet been determined~\cite{Bro1}. The study of the
QGP and of the phase transition is important for understanding
the early evolution of our universe. 
It has been proposed that collisions between highly relativistic
nuclei can be used to recreate conditions of the early universe in
the laboratory. During the past decade,
such experiments have been performed at the BNL AGS and the CERN SPS
at maximum energies ${\rm \sqrt{s_{NN}} =}$4.8 GeV (${\rm Au+Au}$) and
17.3 GeV (${\rm Pb+Pb}$), respectively. The results from those
experiments~\cite{QM93} have demonstrated rich physics which cannot
be explained by a trivial extrapolation of ${\rm p+p}$ results. At the
present time there is, however, no unambiguous evidence for the existence of a
quark-gluon plasma over a significantly large space-time
region. While the analysis of AGS and SPS data
continues, the new Relativistic Heavy Ion Collider (RHIC)
at BNL delivered the first collisions of gold nuclei in June 2000 at center of mass
energies several times larger than those previously available.
The goal of the RHIC heavy ion program is
to study the behavior of strongly interacting matter under conditions
of extreme temperatures and energy densities which are the
prerequisites predicted for the creation of the QGP. To study this 
behavior, the PHOBOS experiment is designed to combine
measurements of global properties, such as the charged particle
multiplicity, with detailed studies aimed at characterizing the
microscopic aspects of the collisions. 
In this paper we present an overview of recent
results from the PHOBOS experiment which provide a coherent picture of the initial
state parton density, which will be reflected by the 
charged particle multiplicities for Au+Au collisions
at different energies. 
Away from the mid-rapidity region, PHOBOS
measurements confirm the hypothesis of limiting
fragmentation in ${\rm Au + Au}$ collisions as function
of centrality. In combination with results from lower
energies and from ${\rm \bar{p} + p}$ collisions, these data permit 
a systematic analysis of particle production mechanisms in nucleus-nucleus~collisions. 
\vspace*{-0.4cm}    
\section{Experimental Setup}\label{sec:setup}
\vspace*{-0.2cm}
The PHOBOS experiment consists of five detector
subsystems. These include a large solid-angle multiplicity array,
vertex finding detectors, two multiparticle tracking spectrometers, a
set of plastic scintillator time-of-flight (TOF) walls, and trigger
detectors. The multiplicity arrays are divided into an octagonal barrel
of silicon pad detectors surrounding the beam pipe in the central
rapidity region, and six ring counters of silicon pad detectors. 
Together, these arrays cover ${\rm \mid\eta\mid\le}$5.4. The vertex detectors consist of two sets of silicon
pad detectors above and below the beam line around the interaction
region covering near mid-rapidity. The spectrometers are positioned on
either side of the
beam partially within a 2T magnet field which provides momentum
measurement and particle identification near mid-rapidity. 
The technical details of the silicon detectors are described~in~Ref~\cite{Nou1}.
Finally, trigger counters
consisting of two rings of 16 ${\rm \check{\rm C}}$erenkov radiators and
two sets of plastic scintillator counters, are arranged around the
beam pipe. They serve as the primary event trigger, and detect charged
particles in the range 4.4 
${\rm <\mid \eta\mid < }$5.2 and 3 ${\rm <\mid \eta\mid < }$ 4.5, respectively. The scintillator counters
are also used for offline event selection and determination of the
collision centrality. Details of this procedure can be found in Ref~\cite{Kat1}.
\vspace*{-0.4cm}     
\section{Initial State Parton Density}
\begin{figure}
\begin{center}
\hspace*{-0.6cm}\psfig{figure=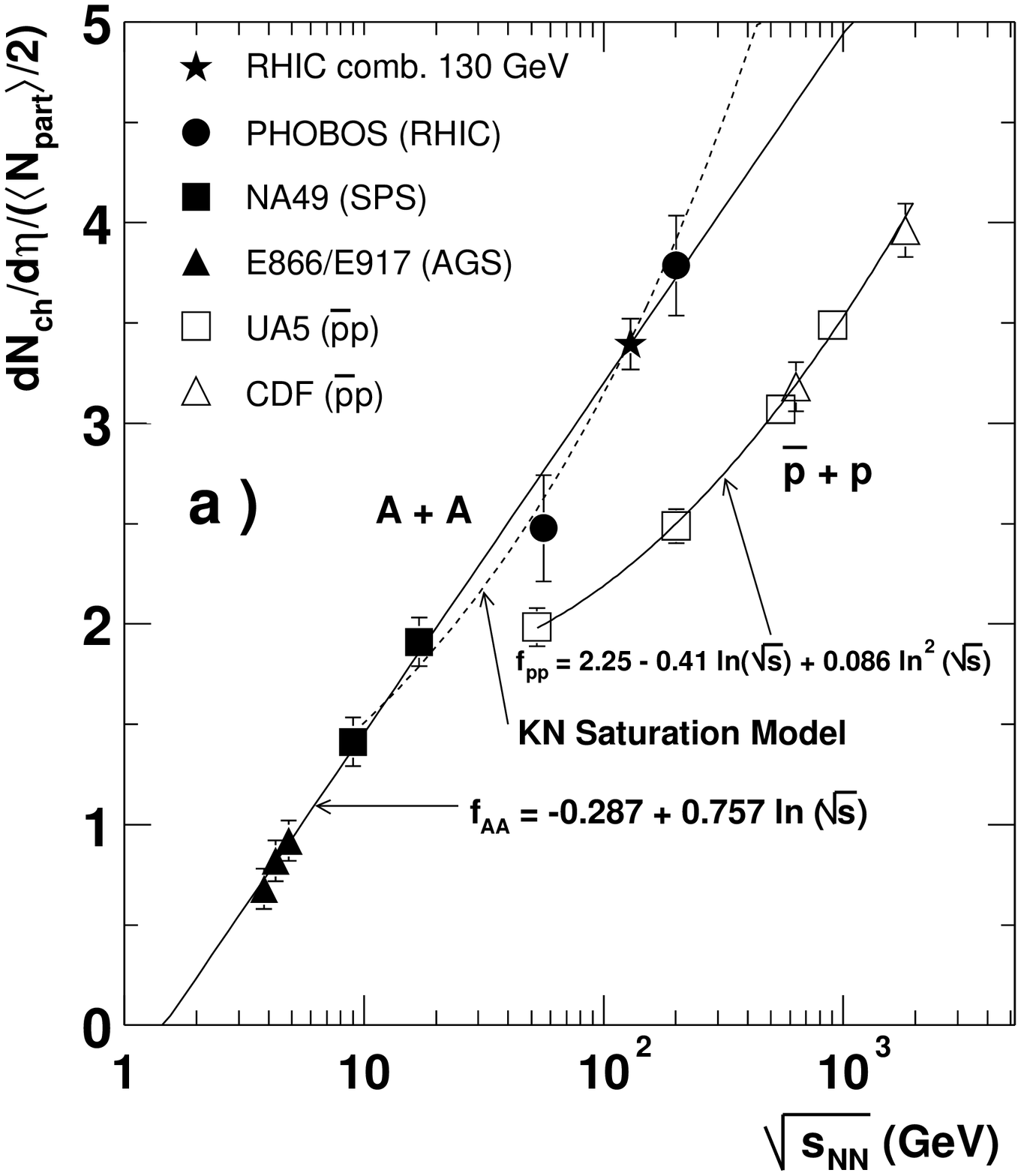,height=2.98in,width=3.1in}
\psfig{figure=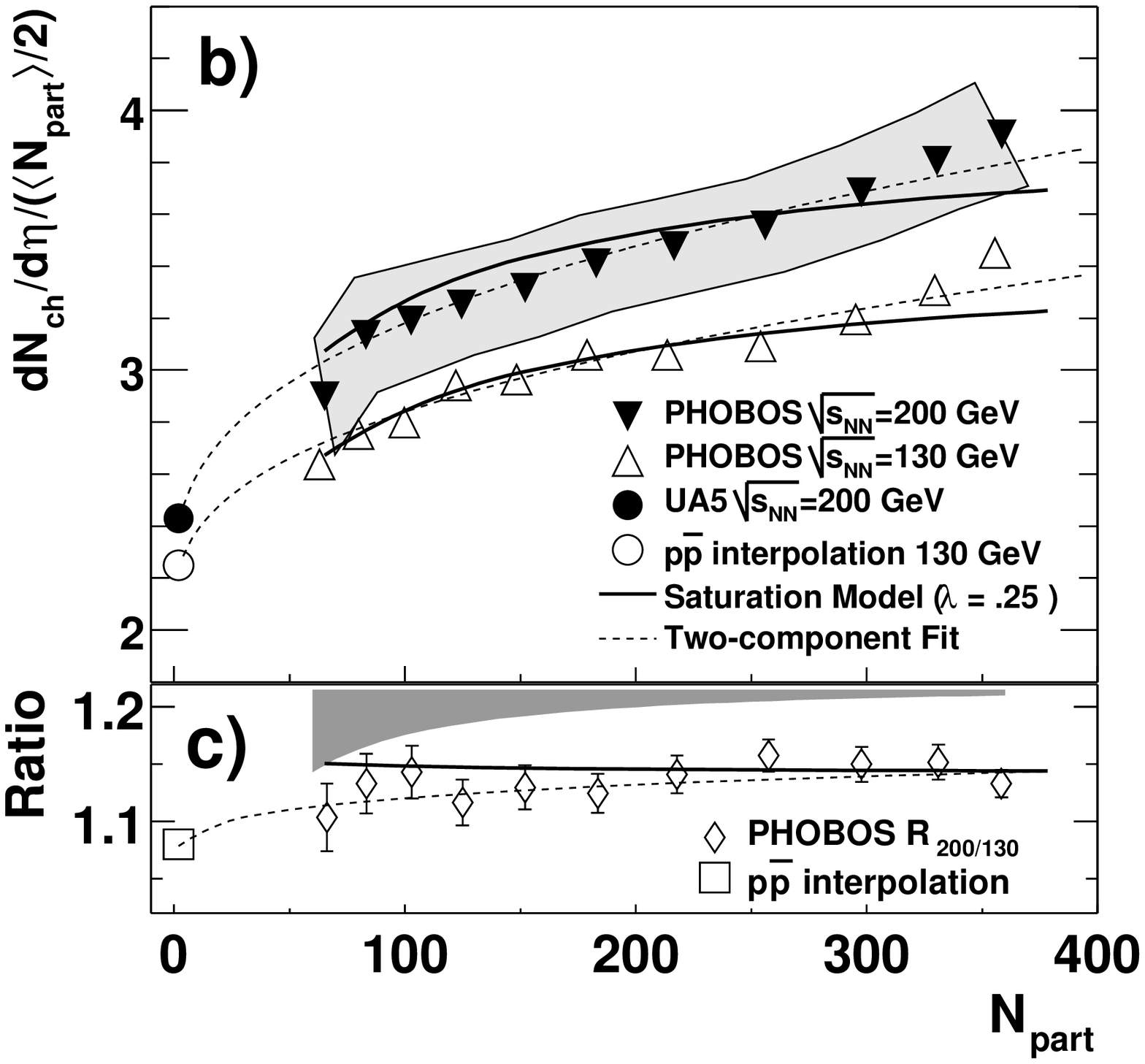,height=2.9in,width=3.1in}
\vspace*{-0.3cm}
\caption{(a) The measured scaled pseudorapidity density ${\rm dN_{ch}/d\eta
/\langle {1\over2}N_{part}\rangle}$
for ${\rm \mid \eta \mid < 1}$ 
in central Au+Au collisions at AGS$^{17}$ and RHIC$^{6}$ and ${\rm Pb+Pb }$
at the SPS${\rm ^{18}}$ (solid points); ${\rm \bar{p}p}$ data from
UA5${\rm ^{8}}$ and
CDF$^{9}$ are shown as open symbols. The continuous lines ${\rm f_{AA}}$ and 
${\rm f_{\bar{p}p}}$ correspond to a fit through the solid points and open
symbols, respectively. (b) The measured scaled pseudorapidity density
as a function of ${\rm N_{part}}$ for Au+Au collisions at 
${\rm \sqrt{s_{NN}} = 130}$ (open triangle) and 200 GeV (closed triangles) 
obtained by PHOBOS. The open and solid circles are ${\rm \bar{p}p}$ results$^{9}$. 
(c) The ratio of charged multiplicities for ${\rm \sqrt{s_{NN}} = }$ 130 and 200 GeV. The gray band indicates the
systematic error estimate. In both panels (b) and (c), results from a
saturation model prediction and two-component fit are shown as solid
and dashed lines, respectively. \ \ \ \ \ \ \ \ \ \ \ \hspace*{11cm}
\label{fig:energy}}
\end{center}
\end{figure}

\vspace*{-0.2cm}     
One of the important observables in heavy-ion interactions is the number of charged particles produced in a
collision and the pseudorapidity density ${ \rm dN_{ch}/d\eta }$. This
number (${ \rm dN_{ch}/d\eta }$) is believed to be proportional to the
entropy density at freeze-out and, since
entropy cannot be destroyed (even in non-equilibrium systems) the
pseudorapidity density provides a constraint on the initial
state parton density and any further entropy produced during the
subsequent evolution. The charged particle
multiplicity at mid-rapidity ${\rm \mid \eta \mid < 1}$ was obtained by
three different analysis procedures~: 1)~Counting of ``tracklets'', i.e. hits in two consecutive Si-detector
planes of the vertex or spectrometer detectors in the field-free
region close to the vertex position. 2)~Counting of hit pads in the
single layer octagon detector, correcting for multiple hits by assuming
Poisson counting statistics. 3)~Relating the energy deposition in the
single-layer octagon detector to the charged-particle
multiplicity. Results from these methods were corrected for
non-vertex background and weak-decay feed-down using a
Geant/HIJING based simulation. The measured scaled pseudorapidity densities
at mid-rapidity for central ${\rm Au + Au}$ collisions are shown in Fig.~\ref{fig:energy}(a) as function of ${\rm \sqrt{s_{NN}}}$. In order
to compare to ${\rm \bar{p}+ p}$ data, ${ \rm dN_{ch}/d\eta }$ is
divided by the number of participant pairs ${\rm \langle
{1\over2}N_{part}\rangle }$.   
A comparison to ${\rm \bar{p}+p}$ collisions~\cite{Ar1}~\cite{Ae2} in the overlapping
energy region shows a ${\rm \sim}$~55~${\rm \%}$ higher production rate in central
${\rm Au + Au }$ collisions, indicating that the ${\rm Au+Au}$ is
not a simple superposition of ${\rm \bar{p}+p}$. 
When compared with results from lower energies for ${\rm Au+Au}$, we observe that    
 ${\rm dN_{ch}/d\eta/\langle{1\over2}N_{part}\rangle}$ for the most central events
increases logarithmically with energy up to the maximum
RHIC energy, ${\rm \sqrt{s_{NN}} = 200 }$ GeV, as indicated by the
solid line fit ${\rm f_{AA} = -0.287 + 0.757 ln
(\sqrt{s}) }$. 
To put these results in context, we also show the prediction of the
parton saturation model (PSM)~\cite{kh1} as a function of energy,
indicated by the dashed line in
Fig.~\ref{fig:energy}(a). 
The good agreement of the PSM with the data suggest that the initial
conditions of the model may be correct for particle
production at the RHIC energies and that these may already be
appropriate at the highest SPS energy.
It is also interesting to extrapolate the prediction of the PSM
and the fit (${\rm f_{AA}}$) to the LHC energy ${\rm \sqrt{s_{NN}} = 5500}$~GeV for the most central
collision of ${\rm Pb + Pb }$. Assuming that the ${\rm \langle {1\over2} N_{part} \rangle = 208 }$,
the pseudorapidity density will be ${\rm (dN_{ch}/d\eta)^{PSM}_{LHC} =
2200 }$ and ${\rm (dN_{ch}/d\eta)^{f_{AA}}_{LHC} = 1300 }$.
These correspond to factors of 2.8 (PSM)
and 1.7 (${\rm f_{AA}}$) increases in
multiplicity normalized to the number of participants relative to the RHIC energy of ${\rm \sqrt{s_{NN}} = 200 
}$ GeV.
In Fig.~\ref{fig:energy}(b), we analyze the particle density as
a function of centrality, expressed as the number of participating
nucleons~\cite{bk2} (${\rm N_{part}}$). Such an analysis may shed light on the relative importance
of soft versus hard processes of particle production and test the
assumption of gluon saturation expected at RHIC~\cite{Wa1}~\cite{Es1}. 
\begin{figure}
\begin{center}
\hspace*{-1cm}\psfig{figure=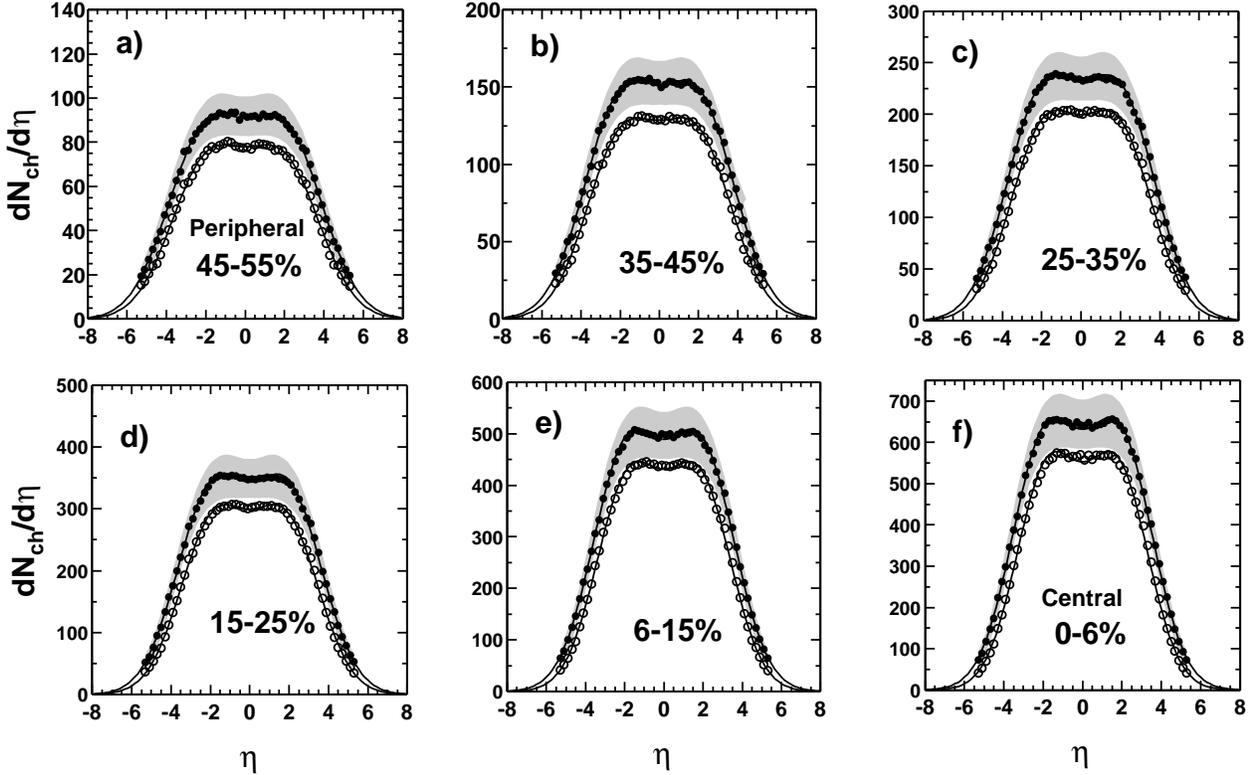,height=4.5in,width=7in}
\vspace*{-0.9cm}
\caption{ Charged particle density distributions are shown for six
centrality bins for ${\rm Au + Au }$ collisions at ${\rm \sqrt{s_{NN}}~= }$130~GeV (open symbols) and 200 GeV (solid points). The continuous line
corresponds to the fit using the stochastic approach in Ref$^{14}$.\hspace*{11.8cm} 
\label{fig:dist}}
\end{center}
\end{figure}
We present the results for ${\rm dN_{ch}/d\eta /\langle {1\over2}N_{part}\rangle}$ vs ${\rm N_{part}}$ for ${\rm \sqrt{s_{NN}}}$ =
200 and 130 GeV. For both energies, we observe a continuous rise of particle
density with increasing centrality. The errors for 200 GeV data are
shown as a gray band and they include both statistical and systematic
errors (90$\%$ C.L.). In order to put these results in context, we  
compare the predictions of PSM, indicated by solid lines and calculations involving fits to
the data using the two-component parametrization proposed in
Ref~\cite{Kh2} (dashed line), 
${\rm dN_{ch}/ d\eta = \big{[}1-x(s)\big{]}n_{pp} {\langle N_{part}\rangle /2}+  x(s)n_{pp} \langle N_{coll} \rangle}$,
where ${\rm x(s)}$ is the fraction of production of charged particles
from hard scattering processes. The energy dependence of
${\rm x(s)}$ has been derived from deep-inelastic ${\rm ep}$ scattering data.
The HERA data have been found to scale as ${\rm x(s) \approx s^{\lambda} }$ with
${\rm \lambda  \approx }$ 0.25 and the ${\rm n_{pp}}$ is the number of particles
produced in single ${\rm p+p}$ collisions where ${\rm N_{part} = 2}$ and
${\rm N_{coll} =1 }$. The number of hard scattering collisions is found to
be ${\rm \langle N_{coll}\rangle = 0.352\langle N_{part}\rangle^{1.37}}$ based on
Ref~\cite{Kh2}. We have performed fits using the measured values for
${\rm n_{pp}}$ and find that ${\rm x(130) = 0.09 \pm 0.02 }$ and ${\rm
x(200) = 0.11 \pm 0.02 }$. According to the assumptions of this
parametrization, the increase in the multiplicity of charged particles
produced from 130 to 200 GeV in ${\rm Au + Au}$ at mid-rapidity
region from hard processes can be obtained by using the values
of ${\rm x(130)}$ and ${\rm x(200) }$ which leads to :      
\begin{equation}
{\rm
{\Bigg{(} { dN_{ch}\over d\eta}\Bigg{)}_{hard}^{200 GeV} \Bigg{/}
\Bigg{(} { dN_{ch} \over d\eta}\Bigg{)}_{hard}^{130 GeV}}  = 
{x(200) n_{pp}(200) <N_{coll}(200)>\over x(130)n_{pp}(130) <N_{coll}(130)>} = 1.34 }
\label{eq:NN}
\end{equation}   
If this description is correct, the increase of 34$\%$ indicates that the hard processes start to play an
significant role at RHIC energies. 
We have also measured ${\rm dN_{ch}/d\eta}$ over the range ${\rm \mid \eta \mid \leq }$ 5.4 for
${\rm Au + Au }$ collisions at ${\rm \sqrt{s_{NN}}= 130}$ and 200 GeV,
which are shown for six centrality bins in Fig.~\ref{fig:dist}. The gray bands represent
the systematic errors (90$\%$ C.L.). The errors at 130 GeV are not shown, but are a
similar in magnitude to those at 200 GeV. A general  feature of these distributions at both
energies and all centrality bins is a flat region extending over about
${\rm \pm 2}$ units of pseudorapidity around mid-rapidity. Outside
this region we observe a fall-off towards large
${\rm\mid\eta\mid \leq 5.4 }$. In Table~\ref{tab:table1}, we present the total charged
particle multiplicity for each centrality bin obtained from the
integration of the data points (${\rm N_{ch}^{data} }$)
and also using a fit (continuous line in Fig.~\ref{fig:dist}) (${\rm N_{ch}^{Fit} }$)
which is based on the stochastic approach~\cite{Bi1}. 
The contribution to the total from the tails of the distributions
outside the measured region is estimated from the
fit to be 2$\%$ for the most central and 4$\%$ for the most peripheral collisions. 
The comparison of ${\rm dN_{ch}/d\eta }$ at 130 and 200 GeV for all centrality
bins shows that additional particle production occurs dominantly in the
mid-rapidity plateau region and also that the width of the distributions is
increased at higher energy.        
\begin{figure}
\psfig{figure=Moriond_figure3_a.eps,height=3.0in,width=3.0in}
\psfig{figure=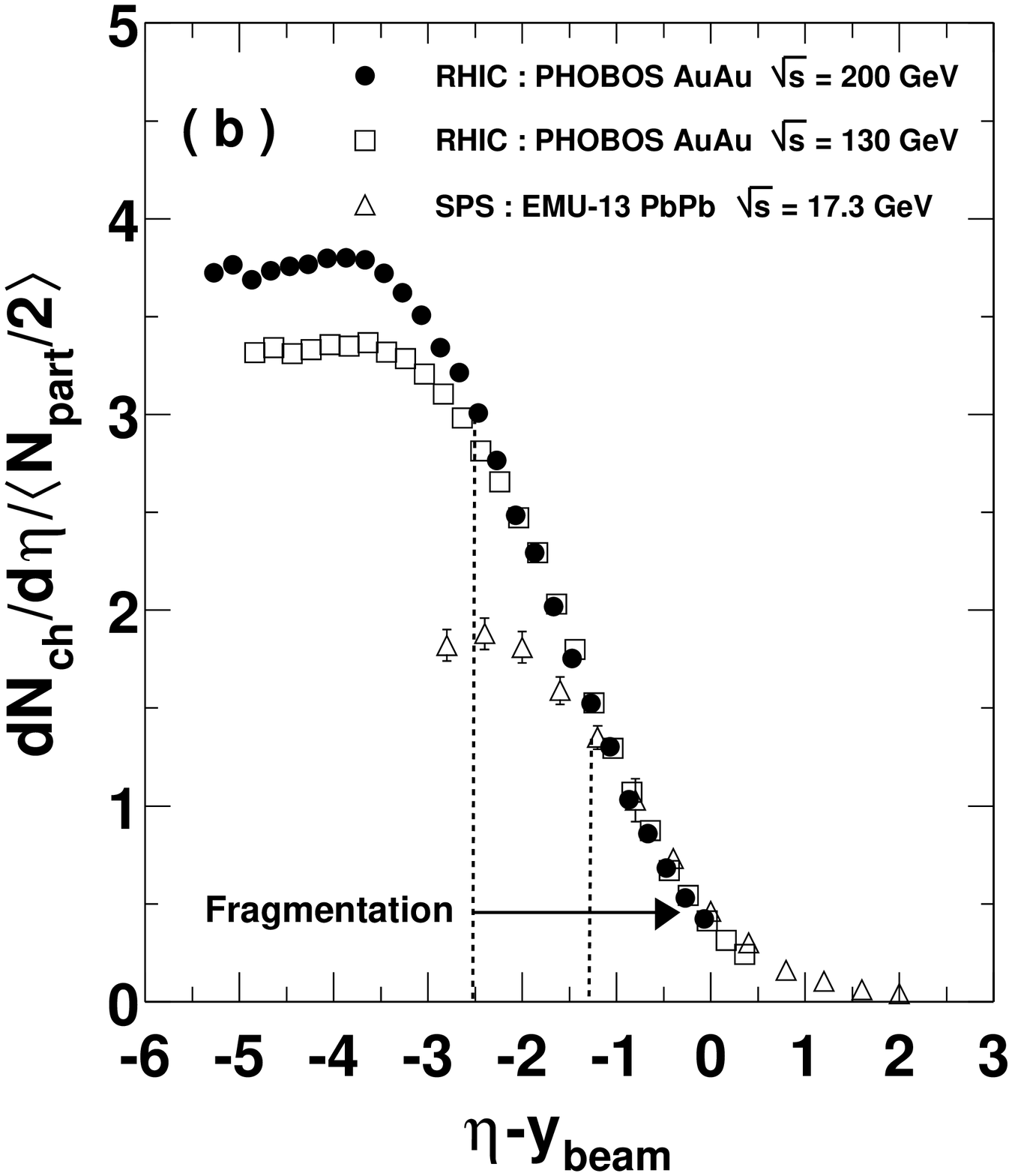,height=3.05in,width=3.0in}
\vspace*{-0.2cm}
\caption{ Illustration of scaling in the fragmentation region for
${\rm \bar{p}+ p}$ (a) and ${\rm Au + Au }$ (b) collisions. \label{fig:AAPP}}
\end{figure}
\vspace*{-0.4cm}
\begin{center}
\begin{figure}
\hspace*{0.5cm}\psfig{figure=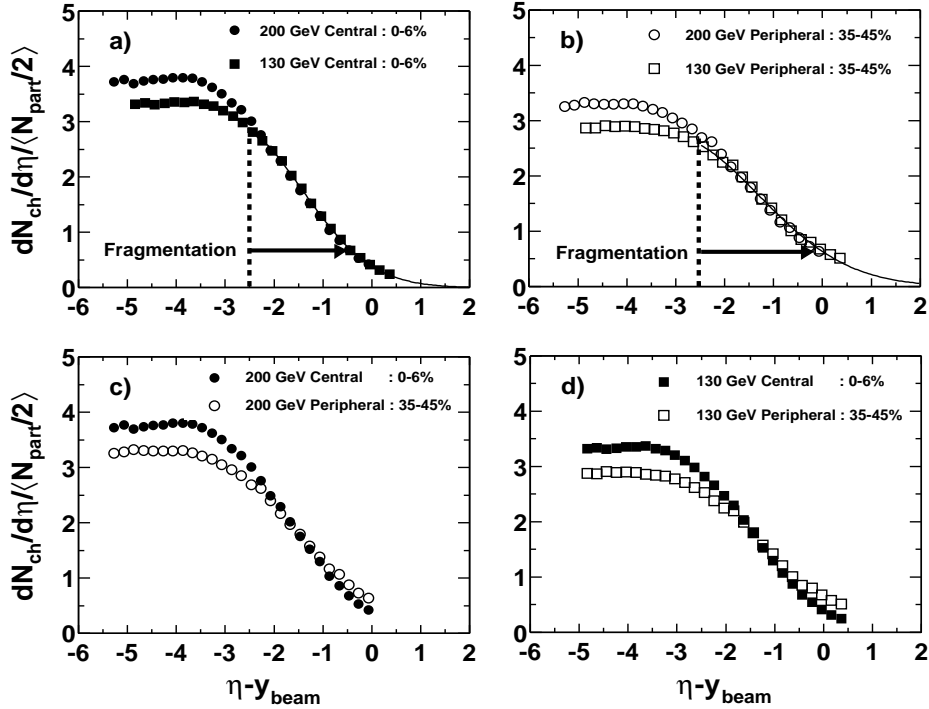,height=3.8in,width=5.3in}
\vspace*{-0.4cm}
\caption{Measured scaled pseudorapidity density ${\rm dN_{ch}/d\eta /\langle{1\over2}N_{part}\rangle}$
vs. ${\rm \eta - y_{beam} }$ for ${\rm Au + Au}$ collisions at ${\rm
\sqrt{s_{NN}}= }$ 130 and 200 GeV for the most central (${6\%}$) and peripheral
(${35-45\%}$). The continuous line corresponds to the fit using the
stochastic approach in Ref$^{14}$, indicating
the limiting fragmentation region.\hspace*{3cm} \label{fig:frag}}
\end{figure}
\end{center}
\vspace*{-0.5cm}
\section{Limiting Fragmentation}
\vspace*{-0.2cm}
In order to understand the relation between particle production and
collision mechanisms, we consider particle production away from
mid-rapidity in the target (or projectile) rest frame. The
distribution of charged particles from ${\rm\bar{p}+p}$ collisions shown in
Fig.~\ref{fig:AAPP}(a) follows a simple scaling relation as predicted by 
``limiting fragmentation'' when plotted versus ${\rm \eta -
y_{beam}}$, where ${\rm y_{beam}}$ is the beam rapidity. 
In Fig.~\ref{fig:AAPP}(b), we observe that the  
${\rm dN_{ch}/d\eta/\langle{1\over 2}N_{part}\rangle}$ distribution for ${\rm Au + Au}$
collisions at 130 and 200 GeV agree for 
 ${\rm \eta -y_{beam} \geq -2.5}$. This indicates that ${\rm Au + Au}$ at 130
GeV also follows limiting fragmentation scaling down to ${\rm \eta -y_{beam} \geq -2.5}$. The comparison between ${\rm Au +
Au}$ at ${\rm \sqrt{s_{NN} }=}$ 130 and 200 GeV and 
${\rm Pb+ Pb}$ at ${\rm \sqrt{s_{NN}} = }$ 17.3 GeV  shows that the extent of the limiting fragmentation
increases by about 1.5 units of pseudorapidity from the maximum SPS energy
to 130 GeV at RHIC. 
In Figs.~\ref{fig:frag}(a) and (b) we show      
${\rm dN_{ch}/d\eta/\langle{1\over 2}N_{part}\rangle}$ vs. ${\rm \eta -y_{beam}}$ for ${\rm Au+Au}$ at
130 and 200 GeV for the most central (6$\%$) and for  
peripheral (${\rm 35-45\%}$) collisions. This analysis has also been performed for all
other centralities between ${\rm 0-45\% }$. We
observe that the extent of the limiting
fragmentation region (${\rm \eta -y_{beam} \geq -2.5 }$) is independent of
centrality. 
In Figs.~\ref{fig:frag}(c) and (d) we show  
${\rm dN_{ch}/d\eta /\langle{1\over2}N_{part}\rangle}$ vs. ${\rm \eta -y_{beam}}$ indicating that
the particle production per participant in the limiting
fragmentation region (${\rm \eta-y_{beam} \geq -1.5 }$) increases from the most
central to the most peripheral at the same energy. 
This increase in particle production near ${\rm \eta-y_{beam} \geq -1.5}$ due to the
target remnants has been previously observed in  ${\rm p + A}$ physics~\cite{Ma1} and
for lower energy ${\rm Pb + Pb}$ collisions~\cite{De1}. The
distribution may also narrow in central collisions due to dynamical
effects, such as baryon stopping, or kinematic effects, such as a shift
in ${\rm \eta - y_{beam}}$ due to the particle mix (${\rm p/\pi}$ ratio) changing
with centrality.
\par
In summary, the first measurements of the pseudorapidity distributions from
${\rm Au + Au}$ collisions at maximum RHIC energy ${\rm \sqrt{s_{NN}}}$
= 200 GeV have been presented and compared to the results
from the new analysis of ${\rm Au + Au}$ data at ${\rm \sqrt{s_{NN} }
= }$ 130~GeV. 
The pseudorapidity distributions have been measured as a function
of collision centrality and energy over an extended range of
pseudorapidity ${\rm \mid \eta \mid \le 5.4 }$, which allows for an
accurate estimate of the total charged particle yield. For the
most central collisions (${\rm 6\%}$) at 130 GeV about 4160 charged particles are
produced reaching about 5050 at the maximum RHIC energy of 200
GeV. The pseudorapidity distributions at 130 and 200 GeV, for all
centrality bins, show evidence that additional particle production
occurs in the mid-rapidity plateau region and that the width of the
distribution increases with energy. Away from mid-rapidity,
a detailed study of the distribution as a function of collision centrality and energy has been
made in order to study limiting fragmentation behavior in ${\rm
Au+Au}$ collisions at RHIC energies.       
A comparison between Au $+$ Au data at ${\rm \sqrt{s_{NN}} =}$~130~GeV and 
${\rm Pb + Pb }$ at SPS energy ${\rm \sqrt{s_{NN}} = }$ 17.3 GeV
indicates that the extent of the limiting fragmentation region grows with
energy by about 1.5 units of ${\rm \eta - y_{beam}}$. We observe also
that the extent of limiting
fragmentation is independent of centrality at the same energy but the
particle production per participant in the limiting fragmentation region increases at high ${\rm \eta -
y_{beam}} $ for more peripheral collisions. This indicates that
the number of charged particles produced grows faster than the
number of participating target nucleons, suggesting that the
``spectator'' matter also contributes to particle production in the
target fragmentation region. 
\begin{table}[t]
\caption{Total charged
particle multiplicity for each centrality bin obtained from the
integration of the data points (${\rm N_{ch}^{data}(\mid\eta\mid\le 5.4 }$))
and also using the fit based on the stochastic
approach$^{14}$(${\rm N_{ch}^{Fit}(all \eta)}$)(see Fig.~\ref{fig:dist}).\ \ 
\label{tab:table1} } 
\begin{center}
\begin{tabular}{|c|ccc|ccc|}
\hline \hline
 &\multicolumn{3}{c|}{200 GeV}&\multicolumn{3}{c|}{130 GeV} \\ 
Bin(\%) & ${\rm N_{ch}^{data}(\mid \eta\mid \le 5.4)}$ & ${\rm
 N_{ch}^{Fit}(all\ \eta) }$ & 
${\rm N_{ch}^{Fit}}$/${\rm N_{ch}^{Data}}$ & 
${\rm N_{ch}^{data}(\mid \eta\mid \le 5.4)}$ & ${\rm N_{ch}^{Fit}(all\ \eta)}$ &
${\rm N_{ch}^{Fit}}$/${\rm N_{ch}^{Data}}$ \\
\hline 
 0 -  6 &4960$\pm$250&5050 &1.02&4100$\pm$210&4160 &1.01\\ 
 6 - 15 &3860$\pm$190&3960 &1.02&3230$\pm$160&3280 &1.01\\ 
15 - 25 &2750$\pm$140&2820 &1.03&2270$\pm$110&2310 &1.02\\ 
25 - 35 &1870$\pm$ 90&1930 &1.03&1540$\pm$ 80&1580 &1.02\\ 
35 - 45 &1230$\pm$ 60&1270 &1.04&1000$\pm$ 50&1030 &1.03\\  
45 - 55 &\ 750$\pm$ 40&\ 780 &1.04&\ 620$\pm$ 30&\ 640 &1.03\\ 
\hline \hline
\end{tabular}
\end{center}
\end{table}
\vspace*{-0.5cm}
\section*{Acknowledgments} 
\vspace*{-0.4cm}
This work was partially supported by U.S. DOE grants DE-AC02-98CH10886,
DE-FG02-93ER40802, DE-FC02-94ER40818, DE-FG02-94ER40865, DE-FG02-99ER41099, and
W-31-109-ENG-38, NSF grants 9603486, 9722606 and 0072204, (Poland) KBN
grant 2 PO3B 04916, (Taiwan) NSC 89-2112-M-008-024.

\end{document}